\documentclass[conference]{IEEEtran}
\IEEEoverridecommandlockouts

\usepackage{cite}
\usepackage{amsmath,amssymb,amsfonts}
\usepackage{algorithmic}
\usepackage{graphicx}
\usepackage{tabularx}
\usepackage{textcomp}
\usepackage{xcolor}
\def\BibTeX{{\rm B\kern-.05em{\sc i\kern-.025em b}\kern-.08em
    T\kern-.1667em\lower.7ex\hbox{E}\kern-.125emX}}
\begin{document}

\title{Comprehensive Study on Lumbar Disc Segmentation Techniques Using MRI Data\\}

\author{\IEEEauthorblockN{1\textsuperscript{st} Serkan Salturk}
\IEEEauthorblockA{\textit{Department of Informatics} \\
\textit{Yildiz Technical University}\\
Istanbul, Turkey \\
ssalturk@yildiz.edu.tr\\
0000-0003-3555-3687}
\and
\IEEEauthorblockN{2\textsuperscript{nd} Irem Sayin}
\IEEEauthorblockA{\textit{Department of Mechatronics} \\
\textit{Yildiz Technical University}\\
Istanbul, Turkey \\
irem.sayin@std.yildiz.edu.tr\\
0000-0002-0627-8308}
\and
\IEEEauthorblockN{3\textsuperscript{rd} Ibrahim Cem Balci}
\IEEEauthorblockA{\textit{Department of Mechatronics} \\
\textit{Yildiz Technical University}\\
Istanbul, Turkey \\
icbalci@yildiz.edu.tr\\
0000-0003-0577-4278}
\and
\IEEEauthorblockN{4\textsuperscript{th} Taha Emre Pamukcu}
\IEEEauthorblockA{\textit{Department of Electronics and Communication} \\
\textit{Yildiz Technical University}\\
Istanbul, Turkey \\
emre.pamukcu@std.yildiz.edu.tr\\
0009-0004-1984-1683}
\and
\IEEEauthorblockN{5\textsuperscript{th} Zafer Soydan}
\IEEEauthorblockA{\textit{Department of Mechatronics} \\
\textit{Nisantasi university}\\
Istanbul, Turkey \\
zsoydan@gmail.com\\
0000-0001-6387-8628}
\and
\IEEEauthorblockN{6\textsuperscript{th} Huseyin Uvet}
\IEEEauthorblockA{\textit{Department of Mechatronics} \\
\textit{Yildiz Technical University}\\
Istanbul, Turkey \\
huvet@yildiz.edu.tr\\
0000-0003-0392-982X}
}

\maketitle

\begin{abstract}
Lumbar disk segmentation is essential for diagnosing and curing spinal disorders by enabling precise detection of disk boundaries in medical imaging. The advent of deep learning has resulted in the development of many segmentation methods, offering differing levels of accuracy and effectiveness. This study assesses the effectiveness of several sophisticated deep learning architectures, including ResUnext, Ef3 Net, UNet, and TransUNet, for lumbar disk segmentation, highlighting key metrics like as Pixel Accuracy, Mean Intersection over Union (Mean IoU), and Dice Coefficient. The findings indicate that ResUnext achieved the highest segmentation accuracy, with a Pixel Accuracy of 0.9492 and a Dice Coefficient of 0.8425, with TransUNet following closely after. Filtering techniques somewhat enhanced the performance of most models, particularly Dense UNet, improving stability and segmentation quality. The findings underscore the efficacy of these models in lumbar disk segmentation and highlight potential areas for improvement.
\end{abstract}


\section{Introduction}

Lumbar disk segmentation is a critical task in medical imaging, aiding in the diagnosis and treatment of spinal disorders. With the advent of deep learning, numerous segmentation techniques have emerged, providing varying degrees of accuracy and efficiency. 
The segmentation of lumbar disks involves the precise delineation of disk boundaries within medical images, typically magnetic resonance imaging (MRI) or computerized tomography (CT) scans. Accurate segmentation is essential for diagnosing conditions such as herniated disks, degenerative disk disease, and spinal stenosis. Traditional methods relied heavily on manual segmentation, which is time-consuming and subject to human error. With advancements in machine learning and neural networks, automated segmentation methods have become increasingly prominent.
This study focuses on comparing the effectiveness of different segmentation methods. By evaluating these methods, we aim to identify the strengths and limitations of each approach in the context of lumbar disk segmentation.
\begin{figure*}[ht]
    \centering
    \includegraphics[width=1\textwidth]{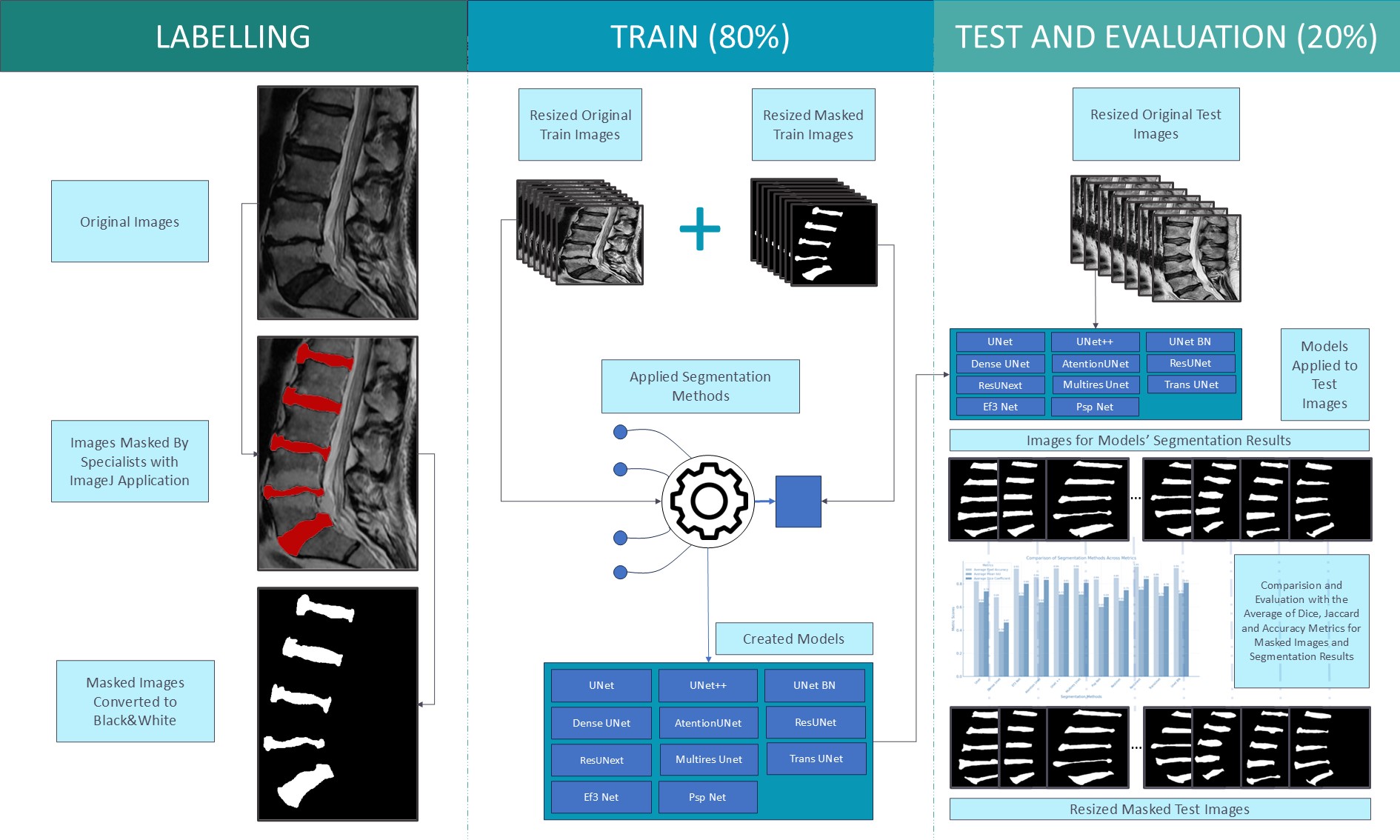} 
    \caption{Graphical Abstract}
    \label{graphical-abstract}
\end{figure*}
\section{Literature Review}

Lumbar intervertebral disc degeneration is a significant cause of lower back pain and disability, necessitating accurate assessment for diagnosis and treatment. The Pfirrmann grading system, evaluating disc degeneration based on MRI signal intensity, structure, and height reduction, is widely used. Recent advances in imaging and deep learning techniques have notably improved the precision and efficiency of lumbar disc segmentation and grading.
An automated 3D lumbar intervertebral disc segmentation strategy from MRI data, utilizing a graphical model-based approach, begins with two user-supplied landmarks. This method extracts the geometrical parameters of all lumbar vertebral bodies and discs from a mid-sagittal slice. Subsequently, a 3D variable-radius soft tube model of the lumbar spine column guides the 3D disc segmentation through multi-kernel diffeomorphic registration between a 3D template of the disc and the observed MRI data. Experiments on 15 patient datasets demonstrated the robustness and accuracy of this algorithm, showcasing significant improvements in segmentation precision \cite{dong2016automated}.
A diagnostic system utilizing T2-weighted sagittal MR images has been developed to diagnose degenerative discs. This system employs a fully automated Expectation-Maximization (EM)-based intervertebral discs (IVD) segmentation technique to segment the lumbar IVD from mid-sagittal MR images. Hybrid features, including basic intensity, invariant moments, and Gabor features, are extracted from the segmented IVDs and classified as degenerative or non-degenerative using a Support Vector Machine (SVM) classifier. Evaluated on 93 clinical sagittal MR images, this system achieved an accuracy of 92.47\%, outperforming other classifiers like k nearest neighbour (kNN) and decision trees, and can serve as a second opinion in diagnosing degenerative discs \cite{beulah2022degenerative}.
A region-based segmentation approach using a region growing algorithm was proposed to segment the lumbar spinal cord from T2-weighted sagittal MRI of the lumbar spine. This method, involving image preprocessing and threshold application to obtain a binary image followed by region growing algorithm, facilitates the detection and analysis of spinal cord diseases \cite{beulah2020spinal}.
Lumbar Spinal Stenosis (LSS) diagnosis benefits from a 3-dimensional automatic segmentation model known as LSS-net. This model performs 3D segmentation on T2 sequence lumbar MR images to diagnose LSS by creating six classes for segmentation, including the spinal disc, canal, thecal sac, posterior element, other regions, and background. The high accuracy of this model, measured by the Intersection over Union (IoU) metric, indicates its potential for creating a Computer Aided Diagnosis system for LSS \cite{altun2023lss}.
An automatic system based on deep convolutional neural networks (CNN) for lumbar disc classification using axial view MRI employs a UNet architecture to localize and detail the herniation location. Utilizing the VGG16 architecture, the system achieved a classification accuracy of 94\%, aiding radiologists in diagnosing and treating lumbar herniated disc disease \cite{mbarki2020lumbar}.
Spine Explorer (Tulong), a deep-learning-based program, automates the acquisition of quantitative measurements for major lumbar spine components on axial lumbar MRIs. This program reduces manual segmentation time and improves measurement accuracy for paraspinal muscles, the disc, and the spinal canal. Spine Explorer demonstrated high intersection-over-union scores and good agreement with manual measurements, supporting its use in clinical practice \cite{shen2021deep}.
A manually segmented lumbar spine MRI database was created to address challenges in robust and accurate segmentation due to varying MRI acquisition characteristics from different sites. This database, including segmentations of lumbar vertebral bodies and intervertebral discs, provides a valuable resource for developing and testing automated segmentation algorithms in multi-domain scenarios \cite{khalil2022multi}.
A method for automatic lumbar vertebrae segmentation in CT images using deep learning involves lumbar spine localization using a UNet network and segmentation using a three-dimensional XUNet method. Validated on public and hospital datasets, this method demonstrated good segmentation performance and potential applications in detecting spinal anomalies and surgical planning \cite{lu2023lumbar}.
The robustness of CNNs for lumbar disc shape reconstruction from MR images was studied, focusing on adversarial robustness to in-distribution (IND) and out-of-distribution (OOD) adversarial attacks. The study found that IND adversarial training improves CNN robustness to adversarial attacks, but defending against OOD attacks remains challenging \cite{chen2021adversarial}.
A CNN model was developed for segmenting and classifying intervertebral disc degeneration (IVDD). The model demonstrated high accuracy and reliability in segmentation and classification, with significant positive impacts on doctors' decision-making when used as an assistive tool (An Automatized Deep Segmentation and Classification Model for Lumbar Disk Degeneration and Clarification of Its Impact on Clinical Decisions).
A method for localizing and automatically segmenting lumbar IVD in 3D from MRI supports finite element (FE) modeling. This method distinguishes between annulus fibrosus (AF) and nucleus pulposus (NP), as well as detects degenerated IVDs, providing accurate and personalized information for clinical applications \cite{matos2023lumbar}.
Several studies have demonstrated the potential of deep learning models in automating the segmentation and grading of lumbar intervertebral discs. CNNs were used for the automatic segmentation and detection of lumbar disc degenerative disease and fractures in MRI images. The study achieved high accuracy in segmenting intervertebral discs and vertebral bodies, but further improvements are needed for detecting degenerative disc disease and fractures \cite{matos2023lumbar}.
 Sun et al. (2023) proposed a high-accuracy quantitation method using the BianqueNet semantic segmentation network, which incorporates a self-attention mechanism and deep feature extraction. This approach achieved high precision in segmenting intervertebral disc-related areas and provided quantitative analysis of degeneration parameters such as signal intensity difference, disc height, and disc height index \cite{sun2023deep}.
Similarly, a study published in PLOS ONE (2023) utilized a CNN to segment and grade lumbar intervertebral discs based on T2-weighted MRI images. This method predicted Pfirrmann grades with an accuracy of 95\%, demonstrating the effectiveness of deep learning in enhancing diagnostic accuracy and consistency \cite{natalia2024lumbar}.
The Pfirrmann grading system remains a cornerstone in the evaluation of lumbar disc degeneration. This system classifies discs into five grades based on MRI characteristics, including signal intensity, disc structure, and height reduction \cite{pfirrmann2001magnetic}. Its clinical relevance has been widely recognized in the diagnosis and management of degenerative disc diseases, as highlighted in various studies \cite{soydan2023automatized,soydan2024tracing}. 
Research has also explored the relationship between Modic changes and Pfirrmann grades in patients with lumbar degenerative diseases. A study published in PLOS ONE (2012) analyzed this correlation and found significant associations between Modic type changes and Pfirrmann grades. This study highlighted the importance of considering both Modic changes and Pfirrmann grades in the comprehensive assessment of disc degeneration \cite{LiPeng2012MRI}.

\section{Method}
In this study, we evaluated the effectiveness of various segmentation techniques for lumbar intervertebral discs in MRI images, building upon a previously established dataset. Our goal was to assess how different segmentation methods perform in accurately identifying and segmenting discs from L1-2 to L5-S1. By comparing these techniques, we aim to determine the most suitable approach for reliable disc segmentation, which could enhance the accuracy of subsequent automized classification and clinical assessments. Figure 1 illustrates the graphical abstract of the study, summarizing the key steps and methodologies employed for segmentation and analysis Figure \ref{graphical-abstract}.

\subsection{Data Collection and Labelling}
Building on the previous dataset \cite{soydan2023automatized}, where intervertebral discs from L1-2 to L5-S1 were segmented and classified using the Pfirrmann grading system, we applied multiple segmentation techniques to assess and compare their accuracy and efficacy. This approach allows us to evaluate the potential improvements in segmentation accuracy and reliability, providing insights into the suitability of each method for lumbar disc analysis. The dataset was obtained from a single center and included patients presenting with low back pain. Patients lacking sagittal T2-weighted images, those with metallic lumbar implants causing image artifacts, or scans of insufficient quality were excluded, resulting in a final sample of 363 patients. Analysis focused on the intervertebral discs from L1-2 to L5-S1. Sagittal T2-weighted images were obtained for each patient using a 1.5-T MRI scanner (GE, Signa, 1.5-T) and subsequently anonymized. The images were acquired using a fast spin-echo sequence with repetition times ranging from 2680 to 4900 ms, echo times between 100 and 109 ms, and an echo train length of 17. The imaging field of view was 32 x 32 cm, with a slice thickness of 4 mm. The medical team consists of two orthopaedists, a neurosurgeon, and a radiologist, all with extensive experience (three with over 15 years, one with over 7 years).

\begin{figure*}[ht]
    \centering
    \setlength{\tabcolsep}{10pt}
    \renewcommand{\arraystretch}{1.5}
    \begin{tabular}{>{\centering\arraybackslash}m{0.3\textwidth} >{\centering\arraybackslash}m{0.3\textwidth} >{\centering\arraybackslash}m{0.3\textwidth}}
        \includegraphics[width=0.8\linewidth]{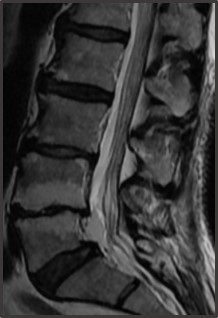} & \includegraphics[width=0.8\linewidth]{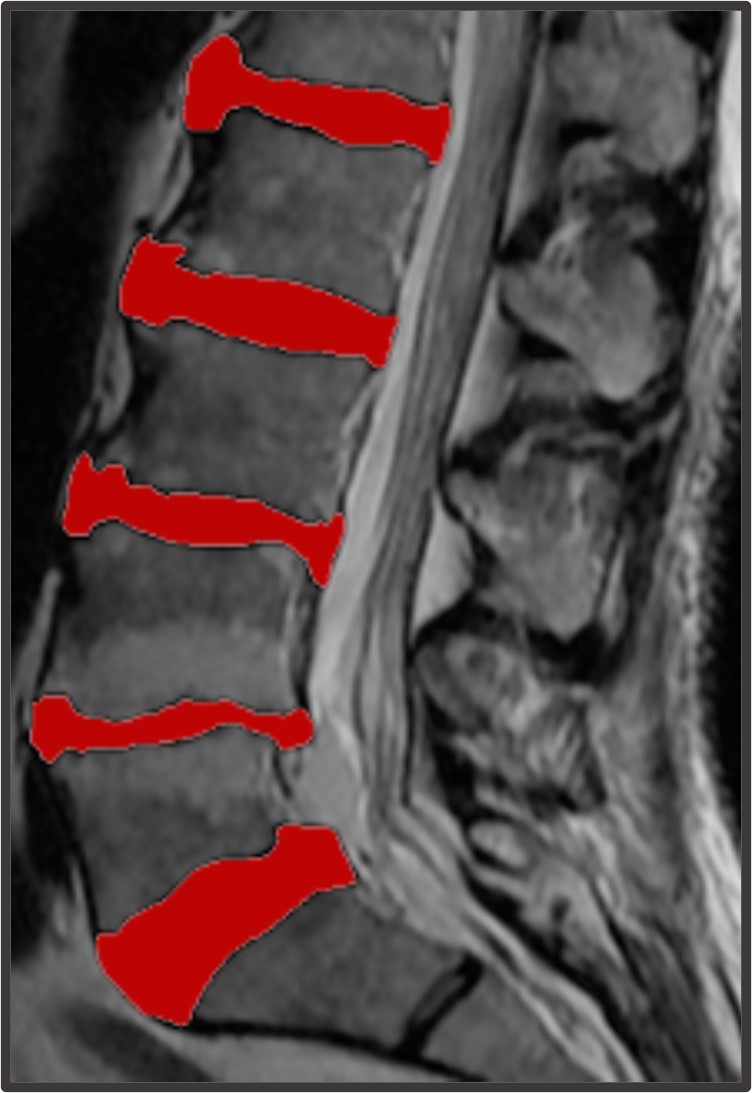} & \includegraphics[width=0.8\linewidth]{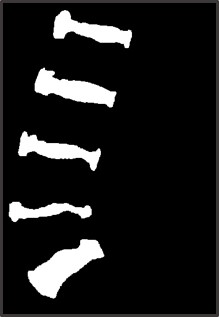} \\
        (a) & (b) & (c) \\
    \end{tabular}
    \caption{(a) Original image (b) Images masked by specialists with ImageJ application (c) Masked images converted to black\&white
}
    \label{data-collection}
\end{figure*}

The imageJ application was used by medical professionals in order to classify lumbar disc images that were collected from patients. All disc regions were marked in white during the labeling procedure, which was subsequently approved by three physicians. Within the context of deep learning algorithms, these data were used as input data Figure \ref{data-collection}.

\subsection{Methodological Framework}

\subsubsection{UNet}

The UNet is a CNN architecture specifically designed for image segmentation, where the objective is to classify each pixel in an image \cite{Ronneberger2015UNet}.

It features a U-shaped structure with two main components: a contracting path (encoder) and an expanding path (decoder). The encoder captures contextual information through downsampling, while the decoder reconstructs the segmentation map through upsampling. Skip connections link corresponding layers in the encoder and decoder, ensuring the preservation of spatial information and facilitating the combination of low-level and high-level features.

The encoder consists of four stages, each with Conv2D layers that progressively increase the number of filters (16, 32, 64, 128), interspersed with Dropout and MaxPooling2D layers for feature extraction and downsampling. At the bottleneck, a block with 256 filters captures high-level features from the downsampled data.

The decoder employs Conv2DTranspose layers to upsample the feature maps, restoring the original resolution. Skip connections from the encoder are concatenated to the upsampled layers to enhance spatial precision.

The model is optimized using the Adam optimizer and is compiled with a custom dice\_loss function and dice\_coefficient metric, both tailored to measure segmentation accuracy.

\subsubsection{UNet++}
UNet++ incorporates additional convolutional layers between the encoder and decoder pathways, creating denser skip connections. These layers refine feature mappings, enhancing segmentation performance by reducing the semantic gap between the encoder and decoder features \cite{zhou2018UNetpp}.

In the implemented UNet++ model, convolutional layers were added to the skip connections of the UNet architecture, while other components of the model remained unchanged.

\subsubsection{UNet BN}
The UNet with batch normalization (BN) is an enhanced version of the original UNet architecture, widely used for image segmentation tasks. In this variant, batch normalization layers are incorporated after convolutional layers to normalize feature maps, which stabilizes and accelerates the training process by minimizing internal covariate shifts. This enhancement improves convergence speed, allows for higher learning rates, and reduces the risk of overfitting. The encoder-decoder structure of UNet, combined with skip connections, facilitates effective learning of both high-level and fine-grained features, while batch normalization further boosts the model's robustness and performance across various segmentation tasks.
In the implemented UNet with batch normalization, BatchNormalization layers were added after each Conv2D layer in the encoder blocks and each Conv2DTranspose layer in the decoder blocks of the original UNet architecture.

\subsubsection{Dense UNet}
In Dense UNet, each layer within a block is directly connected to all subsequent layers in a feed-forward manner. This is achieved by concatenating the output of all preceding layers within the block as input to the current layer \cite{cai2020dense}.

This dense connectivity allows each layer to access feature maps from all prior layers, promoting better feature reuse and reducing the risk of the vanishing gradient problem.

In the Dense UNet implementation, Batch Normalization and ReLU layers were added before each Conv2D layer, and layers within each block were connected using concatenation, based on the original UNet architecture.

\subsubsection{Attention UNet}
The attention block enhances the model's ability to focus on relevant spatial features by selectively weighting different regions of the input tensor. It achieves this by transforming the input feature map $x$ and the gating signal $g$ into a lower-dimensional space using 1x1 convolutions. The transformed inputs are then combined through addition, followed by a ReLU activation, another 1x1 convolution, and a sigmoid activation, which generates a spatial attention map. This attention map is applied to the original input $x$ via element-wise multiplication, enabling the model to emphasize important features and suppress irrelevant ones, thus improving feature refinement during the decoding phase \cite{oktay2018attentionUNetlearninglook}.

In the Attention UNet implementation, attention blocks were added to each decoder block after the first Conv2DTranspose layer, with the inter-channel parameter set to match the number of filters in the Conv2DTranspose layer.

\subsubsection{ResUNet}
ResUNet is an enhanced version of the UNet architecture that incorporates residual connections inspired by ResNet. In this design, residual blocks are integrated into both the encoder and decoder paths, enabling the network to learn identity mappings alongside the standard transformations. These residual connections address the vanishing gradient problem and improve gradient flow during backpropagation, resulting in more efficient training and the capacity to learn deeper representations. By combining UNet's encoder-decoder structure with skip connections and ResNet's residual learning, ResUNet effectively captures both low-level and high-level features, making it particularly suitable for tasks such as medical image segmentation and other pixel-wise prediction problems \cite{Diakogiannis_2020}.

The residual block operates by taking an input tensor and applying two Conv2D layers with ReLU activations and the specified number of filters, along with a dropout layer after the first convolution. Simultaneously, the input tensor is passed through a 1x1 convolution to align its dimensions with the output of the residual block. The final output is obtained by summing the transformed tensor from the convolution operations with the original input tensor, forming the "residual" connection. In the ResUNet model, these residual blocks are employed in both the encoder and decoder paths, following each convolutional layer.

\subsubsection{ResUNext}
The res\_block function in this architecture is designed to integrate residual connections with a focus on efficient gradient propagation and feature refinement. Unlike traditional blocks, it incorporates two convolutional layers, each followed by batch normalization and ReLU activation, with a shortcut connection directly linking the input to the output. This structure emphasizes identity mapping and smooth feature learning, essential for stable training and deeper architectures \cite{balasundaram2024novel}.

In this model, residual blocks play a dual role. In the encoder, they enhance feature extraction by maintaining the flow of critical information across layers while max-pooling reduces spatial dimensions. At the bottleneck, the residual block operates with the maximum filter count, ensuring that the most abstract and high-level features are effectively captured. In the decoder, the blocks refine features after transposed convolutions and skip connections, merging spatial and contextual information for precise reconstructions. This approach enables ResUNet to maintain a balance between detail preservation and high-level abstraction, making it robust for tasks requiring fine-grained predictions, such as segmentation.

\subsubsection{Multires UNet}
This multi-resolution UNet model employs a tailored architecture for image segmentation by integrating convolutional layers with varying kernel sizes, enabling the extraction of information at multiple resolutions. The encoder, or contracting path, uses convolutional layers followed by max-pooling to progressively reduce spatial dimensions while capturing high-level features. At each stage, parallel convolutional layers with kernel sizes of 3x3 and 5x5 are applied, and their outputs are concatenated, allowing the model to capture both detailed and broad features. Dropout layers are incorporated after each convolution to reduce overfitting by randomly deactivating a subset of neurons during training. As the depth of the network increases, the number of filters grows, enabling the extraction of increasingly complex features \cite{Ibtehaz_2020}.

The decoder, or expanding path, mirrors the encoder structure by using transposed convolutions to upsample feature maps and restore spatial resolution. At each stage, the upsampled features are concatenated with the corresponding feature maps from the encoder, leveraging skip connections to recover lost spatial details. This mechanism ensures the retention of crucial fine-grained information from earlier layers, aiding in the precise reconstruction of the segmentation map.

\subsubsection{TransUNet}
TransUNet combines the strengths of transformer-based architectures and CNNs to achieve state-of-the-art performance in image segmentation tasks. Unlike traditional UNet models, TransUNet integrates a transformer module within the encoder, enabling the capture of global contextual information alongside local spatial features. This hybrid design is particularly effective for complex segmentation tasks where both fine-grained details and long-range dependencies are critical \cite{chen2021transUNettransformersmakestrong}.
This architecture is well-suited for tasks requiring precise delineation of regions, such as medical image segmentation. The combination of transformer modules and UNet's hierarchical design allows TransUNet to achieve superior performance by effectively capturing both global context and detailed features.

\subsubsection{EF3 Net}
The architecture builds upon the UNet structure but incorporates an EfficientNet-B3 (EF3) backbone as the feature extractor. This hybrid design begins with the EfficientNet-B3 encoder, which leverages advanced convolutional blocks optimized for both accuracy and efficiency. The EfficientNet-B3 architecture employs mobile inverted bottleneck convolutions (MBConv) and squeeze-and-excitation blocks to enhance feature extraction and reduce computational demands. These components enable the model to capture detailed spatial features while maintaining a low parameter count. After feature extraction, the decoder utilizes transposed convolutions to upsample the feature maps and restore spatial resolution, a key characteristic of UNet’s segmentation framework. Skip connections bridge the encoder and decoder, transferring essential low-level spatial features to ensure accurate segmentation \cite{tan2019efficientnet}.

The main distinction between this model and the original UNet lies in the encoder. While UNet uses a straightforward sequence of convolutional layers, the proposed architecture integrates the more advanced and efficient EfficientNet-B3 backbone. EfficientNet achieves a balanced scaling of depth, width, and resolution, resulting in improved performance with reduced computational cost compared to the simpler UNet encoder. However, both architectures share a similar decoder design that employs upsampling and skip connections, crucial for pixel-level segmentation tasks. The EfficientNet-based approach typically delivers higher accuracy, particularly on complex or large datasets, albeit at the cost of increased implementation complexity and a greater need for fine-tuning compared to the standard UNet.

\subsubsection{Psp Net}
The Pyramid Scene Parsing (PSP) network is an advanced deep learning model designed for semantic segmentation, excelling at capturing contextual information across multiple scales \cite{zhao2017pyramidsceneparsingnetwork}. The architecture employs a convolutional encoder-decoder framework, featuring layers of convolutional operations interspersed with max pooling. This setup enables the network to extract hierarchical features while gradually reducing the spatial dimensions of the input image. At its core, the network utilizes deep convolutional layers in the bottom layer to extract rich feature representations, which are then passed to the Pyramid Pooling Block (PPB). The PPB is a key element that captures global context by applying average pooling at multiple bin sizes, producing feature maps that encode information from different spatial scales. These pooled features are subsequently processed through convolutional layers and upsampled to align with the original input dimensions.

After the pyramid pooling stage, the network transitions into the decoder path, where transposed convolutions are used to upsample the feature maps. These upsampled features are concatenated with corresponding encoder features through skip connections, ensuring the retention of spatial information lost during downsampling. This approach enhances the network’s ability to produce accurate and detailed segmentation maps.

\subsection{Hyperparameter Optimization}

In the optimization of the UNet model with Batch Normalization (BN), grid search was employed to fine-tune key hyperparameters, including batch size, the number of Conv2D filters, and dropout rates. The ResUNext model had previously demonstrated superior performance in earlier experiments, prompting further exploration of its potential for improvement. The grid search method enabled a systematic exploration of various hyperparameter combinations to determine the optimal configuration for enhancing segmentation performance.
 
\subsection{Comparision Metrics}
Pixel accuracy, IoU, and Dice coefficient are widely used metrics for evaluating segmentation models, each highlighting different performance aspects. Pixel accuracy quantifies the percentage of correctly classified pixels across the entire image, offering a general overview of performance but failing to address class imbalance. IoU measures the overlap between predicted and ground truth regions by dividing their intersection by their union, making it effective for assessing the model's ability to capture object shapes and boundaries. The Dice coefficient, also an overlap-based metric, places greater emphasis on the intersection by doubling it, making it particularly sensitive to smaller regions. This sensitivity often makes Dice the metric of choice in medical imaging or detailed segmentation tasks. In summary, pixel accuracy provides an overall performance perspective, while IoU and Dice coefficient offer deeper insights into spatial overlap precision. 

In this study, 10-fold cross-validation was utilized to evaluate model performance. Table \ref{tab1} presents the average performance across all folds for each model, along with the results from the best-performing fold, providing insight into both overall effectiveness and peak accuracy.

To refine the predicted segmentation results, a filter was applied to retain only the five largest connected components in each image, with all smaller regions painted black. This approach focuses on the most significant segments, reducing noise and excluding irrelevant smaller areas that might otherwise affect evaluation. By preserving the top five largest regions, the analysis emphasizes the most meaningful parts of the segmentation while minimizing distortions from less significant fragments.

\section{Results}

This study involves the segmentation of lumbar discs utilizing many sophisticated deep learning models, each assessed against critical parameters to identify the most efficient method for precise segmentation. Models including ResUnext, Ef3 Net, UNet++, Dense UNet, and TransUNet were evaluated and compared according to Average Pixel Accuracy, Mean IoU, and Dice Coefficient.

The results indicate that ResUnext achieved the highest accuracy, with a Pixel Accuracy of 0.9492, a Mean IoU of 0.7505, and a Dice Coefficient of 0.8425, so positioning it as the strongest model in segmentation precision. UNet++ and Ef3 Net exhibited robust segmentation performance, with UNet++ attaining a Pixel Accuracy of 0.9350, Mean IoU of 0.7092, and Dice Coefficient of 0.8100, while Ef3 Net attained a Pixel Accuracy of 0.9321, Mean IoU of 0.7015, and Dice Coefficient of 0.8027. Models such as UNet++ and Multires UNet maintained consistent accuracy regardless of filtering, demonstrating their robustness, while Dense UNet saw marginal improvements as a result of filtering. Table \ref{tab1} shows results of models and filtered models.

These results highlight the superiority of ResUnext as the foremost model for lumbar disk segmentation, with UNet++ and Ef3 Net also demonstrating dependable performance. This approach holds promise for advancing clinical decision-making, facilitating precise diagnosis and treatment planning in spinal healthcare.

\begin{table*}[htbp]
\caption{Model Results}
\renewcommand{\arraystretch}{1.4}
\begin{center}

\begin{tabularx}{\textwidth}{|>{\arraybackslash}m{3cm}|>{\centering\arraybackslash}m{2cm}|>{\centering\arraybackslash}m{2cm}|>{\centering\arraybackslash}m{2cm}|>{\centering\arraybackslash}m{2cm}|>{\centering\arraybackslash}m{2cm}|>{\centering\arraybackslash}m{2.1cm}|}
\hline       
 \textbf{Model}& \textbf{Average Pixel Accuracy}& \textbf{Average Mean IoU}& \textbf{Average Dice Coefficent}& \textbf{Max Pixel Accuracy}& \textbf{Max Mean IoU} & \textbf{Max Dice Coefficent} \\
\hline
\hline
UNet & 0.8489 & 0.6426 & 0.7359 & 0.9377 & 0.7144 & 0.8140 \\
\hline
UNet (Filtered) & 0.8490 & 0.6429 & 0.7361 & 0.9377 & 0.7146 & 0.8142 \\
\hline
UNet++ & 0.9350 & 0.7091 & 0.8099 & 0.9373 & 0.7135 & 0.8132 \\
\hline
UNet++ (Filtered) & 0.9350 & 0.7092 & 0.8100 & 0.9374 & 0.7136 & 0.8133 \\
\hline
UNet BN & 0.9370 & 0.7192 & 0.8102 & 0.9386 & 0.7150 & 0.8144 \\
\hline
UNet BN (Filtered) & 0.9390 & 0.7159 & 0.8151 & 0.9354 & 0.7105 & 0.8094 \\
\hline
Dense UNet  & 0.6858 & 0.3888 & 0.4669 & 0.9319 & 0.5472 & 0.6405 \\
\hline
Dense UNet (Filtered) & 0.6957 & 0.3927 & 0.4679 & 0.9356 & 0.5665 & 0.6599 \\
\hline
Atention UNet & 0.8589 & 0.6416 & 0.8358 & 0.9375 & 0.7138 & 0.8135 \\
\hline
Atention UNet (Filtered) & 0.8589 & 0.6427 & 0.8359 & 0.9376 & 0.7140 & 0.8137 \\
\hline
ResUNet & 0.8529 & 0.6549 & 0.7460 & 0.9439 & 0.9439 & 0.8289 \\
\hline
ResUNet (Filtered) & 0.8530 & 0.6550 & 0.7461 & 0.9441 & 0.7333 & 0.8294 \\
\hline
ResUNext & 0.9491 & 0.7505 & 0.8425 & 0.9512 & 0.7628 & 0.8528 \\
\hline
ResUNext (Filtered) & \textbf{0.9492} & \textbf{0.7505} & \textbf{0.8425} & 0.9512 & 0.7626 & 0.8527 \\
\hline
Multires UNet & 0.9354 & 0.7089 & 0.8095 & 0.9372 & 0.7129 & 0.8128 \\
\hline
Multires UNet (Filtered) & 0.9354 & 0.7089 & 0.8095 & 0.9372 & 0.7128 & 0.8127 \\
\hline
TransUNet & 0.8636 & 0.6976 & 0.7806 & 0.9523 & 0.7748 & 0.8622 \\
\hline
TransUNet (Filtered) & 0.8638 & 0.6982 & 0.7810 & \textbf{0.9526} & \textbf{0.7758} & \textbf{0.8629} \\
\hline
Ef3 Net & 0.9317 & 0.7007 & 0.8021 & 0.9409 & 0.7209 & 0.8192 \\
\hline
Ef3 Net (Filtered) & 0.9321 & 0.7015 & 0.8027 & 0.9409 & 0.7208 & 0.8191 \\
\hline
Psp Net & 0.8393 & 0.6023 & 0.6865 & 0.9465 & 0.7480 & 0.8415 \\
\hline
Psp Net (Filtered) & 0.8482 & 0.6117 & 0.6943 & 0.9466 & 0.7481 & 0.8415 \\
\hline
\end{tabularx}
\label{tab1}
\end{center}
\end{table*}

\section{Conclusion and Future Work}
In this work, multiple deep learning architectures were used to segment lumbar disks, and each approach was assessed using a variety of measures such as Pixel Accuracy, Mean IoU, and Dice coefficient. The ResUnext model had the greatest performance for accuracy and segmentation quality among the studied architectures, with TransUNet closely following. ResUnext attained an Average Pixel Accuracy of 0.9492 and an Average Dice Coefficient of 0.8425, indicating strong and reliable segmentation performance. Filtered iterations of the models produced marginal improvements in accuracy and stability, especially with the Dense UNet, where the filtering process contributed to the improvement of both the Mean IoU and Dice Coefficient.

The evaluation indicates that the selected designs are efficient for lumbar disk segmentation, however there exists potential for enhancement. In future endeavors, supplementary segmentation techniques will be investigated  to improve precision and computing efficacy. Subsequent to the segmentation process, a fully automated classification model will be included to detect and classify certain lumbar disk diseases based on the segmented areas. This multi-phase methodology may enhance the precision and clinical utility of models, hence facilitating diagnosis and therapy planning for spinal disorders.

\section*{Acknowledgment}
The authors extend their gratitude to the Advanced System and Invention Lab (ASILAB) for their support and to Dr. Emru Bayramoglu and Dr. Recep Karasu for their invaluable contributions in annotating the data.

\bibliographystyle{ieeetr}
\bibliography{main}

\end{document}